# CT Radiomics-Based Explainable Machine Learning Model for Accurate Differentiation of Malignant and Benign Endometrial Tumors: A Two-Center Study

Tingrui Zhang[1,2,†], Honglin Wu[3,†], Zekun Jiang[4], Yingying Wang[5], Rui Ye[6], Huiming Ni[1], Chang Liu[4], Jin Cao[4], Xuan Sun[1], Rong Shao[7], Xiaorong Wei[1], Yingchun Sun[1,*]

[1] Gynecology Department, Qingdao Hiser Hospital Affiliated of Qingdao University (Qingdao Traditional Chinese Medicine Hospital), Qingdao 266000, China

[2] Sichuan University - Pittsburgh Institute, Sichuan University, Chengdu 610000, China

[3] Department of Obstetrics and Gynecology, Qingbaijiang Women's and Children's Hospital (Maternal and Child Health Hospital), West China Second University Hospital, Sichuan University, Chengdu 610300, China

[4] College of Computer Science, Sichuan University, Chengdu, Sichuan 610000, China

[5] Radiology Department, Qingdao Hiser Hospital Affiliated of Qingdao University (Qingdao Traditional Chinese Medicine Hospital), Qingdao 266000, China

[6] Department of Traditional Chinese Medicine, Jiaozhou Traditional Chinese Medicine Hospital, Qingdao 266000, China

[7] Adult Traditional Chinese Medicine Department, Qingdao Women and Children's Hospital, Qingdao 266000, China.

[†]These authors contributed equally to this work and share first authorship.

*Corresponding authors: Yingchun Sun. Gynecology Department, Qingdao Hiser Hospital Affiliated of Qingdao University (Qingdao Traditional Chinese Medicine Hospital), Qingdao 266000, China. Email: sunny8223@126.com.

**Abstract**

**Objectives:** This study aimed to develop and validate a CT radiomics-based explainable machine learning model for precise diagnosing malignancy and benignity specifically in endometrial cancer (EC) patients.

**Methods:** A total of 83 EC patients from two centers, including 46 with malignant and 37 with benign conditions, were included, with data split into a training set (n=59) and a testing set (n=24). The regions of interest (ROIs) were manually segmented from pre-surgical CT scans, and 1132 radiomic features were extracted from the pre-surgical CT scans using Pyradiomics. Six explainable machine learning (ML) modeling algorithms were implemented respectively, for determining the optimal radiomics pipeline. The diagnostic performance of the radiomic model was evaluated by using sensitivity, specificity, accuracy, precision, F1 score, area under the receiver operating characteristic curve (AUROC), and area under the precision-recall curve (AUPRC). To enhance clinical understanding and usability, we separately implemented SHAP analysis and feature mapping visualization, and evaluated the calibration curve and decision curve.

**Results:** By comparing six modeling strategies, the Random Forest model emerged as the optimal choice for diagnosing EC, with a training AUROC of 1.00 and a testing AUROC of 0.96. SHAP identified the most important radiomic features, revealing that all selected features were significantly associated with EC ($P < 0.05$). Radiomics feature maps also provide a feasible assessment tool for clinical applications. Decision Curve Analysis (DCA) indicated a higher net benefit for our model compared to the "All" and "None" strategies, suggesting its clinical utility in identifying high-risk cases and reducing unnecessary interventions.

**Conclusion:** CT radiomics-based explainable ML model achieved high diagnostic performance, which could be used as an intelligent auxiliary tool for the diagnosis of endometrial cancer.

**Highlight:**

1. This is the first study to apply CT radiomics-based machine learning for precise diagnosing the malignancy and benignity of endometrial cancer (EC).
2. Among six modeling methods, the Random Forest model, using radiomic features from pre-surgical CT scans, demonstrated the highest diagnostic accuracy for EC.
3. CT scans are widely available, making the development of a CT radiomics-based intelligent diagnostic model highly valuable for accurately identifying tumor characteristics and guiding clinical decision-making in EC.

## Introduction

Endometrial cancer (EC) is the second most common cancer among women [1]. Despite progress in cancer research, the incidence and mortality rates of EC continue to rise [2]. In high-income countries, EC is the most frequently diagnosed gynecologic cancer, and its incidence is steadily increasing worldwide. Since the mid-2000s, the annual incidence of uterine malignancies has risen by approximately 1% in women over 50 and by 2% in younger women since the mid-1990s. By the end of 2023, it is estimated that there will be 66,200 new cases and 13,030 related deaths in the United States alone [3].

Risk factors for EC include prolonged exposure to unopposed estrogen, often associated with conditions like polycystic ovary syndrome or infertility treated with tamoxifen, as well as obesity and hyperinsulinemia [4]. The most common symptom of EC is abnormal uterine bleeding, which should always be evaluated in postmenopausal women or women with other risk factors [5]. Preoperative imaging and histologic assessments are essential for tailoring surgical approaches to avoid unnecessary lymphadenectomy (LND) in low-risk patients [6]. Generally, an abdominopelvic computed tomography (CT) scan is performed to assess lymph node (LN) involvement or distant metastases, and positron emission tomography/CT scans are also viable options. Chest CT scans should be part of the initial assessment to rule out lung metastasis in high-risk cases. The role of serum tumor markers in EC is still uncertain [7].

As radiomic technology has advanced in recent years, increased attention is being applied to its role in managing endometrial cancer [8-9]. Sophisticated quantitative, high-throughput radiomic features can be derived from tumor volumes in medical imaging, allowing for the identification of tumor characteristics that are imperceptible to the human eye. Radiomics provides a powerful method for capturing intratumoral heterogeneity by analyzing features from radiological images, and it has already contributed to the creation of diagnostic, predictive, and prognostic models, advancing the field of personalized medicine [10-13].

Artificial intelligence (AI), especially machine learning (ML), has shown substantial benefits in diagnosing endometrial cancer (EC), particularly in identifying key features such as lymph node metastases and lymphovascular space invasion [14]. These advantages are most evident in studies that employ magnetic resonance imaging (MRI) due to its superior imaging details [15-16]. However, CT scans are more commonly used in routine clinical practice for initial diagnoses because of their faster acquisition times and broader availability. Therefore, it is crucial to develop ML models that can effectively analyze and extract features from CT scans. To date, CT-based radiomics combined with ML models have shown promising results in diagnosing lung cancer [17], skin cancer [18], prostate cancer [19], and colorectal cancer [20], but similar analyses for EC are limited to the pre-identification of MMR-D or TMB-H [21] and to recurrence prediction for prognosis [22].

Therefore, unlike previous studies that have predominantly focused on MRI, our study aims to develop and test an explainable ML model designed to diagnose EC using radiomics data extracted from pre-surgical CT scans. This model is intended to support early diagnosis of EC and facilitate informed treatment decision-making.

## Result

### Clinical characteristics

There were 33 malignant and 26 benign EC cases in the training set, and 13 malignant and 11 benign EC cases in the testing set. We further divided the data into an experimental group (59 cases) and a control group (24 cases) based on the malignancy and benignity of EC for subsequent radiomics feature distribution analysis. **Table 1** lists all the obtained clinical characteristics, with no statistically significant differences between the two groups (P = 0.218–1).

### Radiomics features discovery

**Fig. 1** presents the SHAP analysis and importance ranking of the top 20 radiomics features. Among them, 60% belong to texture features, while 40% are first-order statistical features. Additionally, 90% of these features originate from transformed images, indicating that image transformations such as wavelet and LoG filtering can enhance the expression of texture features, which is consistent with previous studies [23].

**Table 2** presents a comparison of the top 20 radiomic features between experimental and control groups, highlighting significant differences in their means and variability. Specifically, features such as log-sigma-5-0-mm-3D_firstorder_90Percentile and log-sigma-1-0-mm-3D_glcm_MaximumProbability show substantial differences between groups, with the experimental mean significantly higher or lower than the control mean across all features, as indicated by the p-values (all $p < 0.001$). For instance, the feature log-sigma-5-0-mm-3D_firstorder_90Percentile had a mean of 0.90±2.77 in the experimental group compared to -0.97±0.72 in the control group, reflecting a pronounced disparity in their distributions. These statistically significant differences suggest that the radiomic features can potentially distinguish between conditions applied in experimental versus control settings, providing a quantitative basis for further model development in predictive machine learning applications within medical imaging.

**Radiomics feature maps**

The feature maps in **Fig. 2** visually demonstrate the distinct radiomics characteristics between malignant and benign tumors. The malignant tumor (top row) displays greater heterogeneity across multiple feature domains, with more complex internal patterns and irregular intensity distributions. In contrast, the benign tumor (bottom row) exhibits more homogeneous patterns with uniform intensity and smoother transitions. This visualization approach provides an intuitive representation of the complex mathematical features that drive our classification model.

**Development of machine learning models**

The predictive performance comparison of all machine learning models is detailed in **Table 3**, demonstrating that all radiomic models were capable of diagnosing EC with AUROCs exceeding 0.88 in the testing set. However, the Random Forest model was particularly distinguished, achieving an AUROC of 0.96, an AUPRC of 0.94, a sensitivity of 100%, and a specificity of 92.31% in testing set, which was superior to all other models (P<0.05). Additionally, we also conducted 5-fold cross-validation, with the results presented in Supplementary Table 1.

**Overall performance of the final radiomics model**

The performance metrics and evaluation results of the final radiomics model are shown in **Table 3** and **Fig. 3.** In the training set, the Random Forest model achieved a perfect AUROC of 1.00 and AUPRC of 1.00, along with 95.83% specificity and 100% sensitivity, underscoring its excellent diagnostic accuracy. Similarly, in the testing set, this model maintained a high AUROC of 0.96 and AUPRC of 0.94, with specificity of 92.31% and sensitivity of 100%, thereby confirming the effectiveness of our final radiomic signature. Furthermore, the confusion matrices also reveal the high precision of the model (**Fig. 3A, B**), solidifying its role as the most reliable model for EC diagnosis. In addition, the ROC curves for both the training and testing sets of the best-performing model are also plotted (see Supplementary Figure 1), providing a visual confirmation of its robust discriminative ability.

To further evaluate the clinical applicability of our model, we conducted calibration curve analysis based on Isotonic Regression [23] (**Fig. 4A, B**) and decision curve analysis (**Fig. 4C, D**). The calibration curve further assessed the model's true prediction bias, while the decision curve demonstrated that our model consistently achieved a higher net benefit across different risk thresholds compared to the "treat all" or "treat none" strategies. These clinical analyses further highlight the potential clinical value of the radiomics model, providing important decision support for EC precise diagnosis.

**Discussion**

This study developed and validated an explainable ML model for the precise identification of malignant EC, filling the research gap in CT radiomics for EC diagnosis and exploring its potential value in clinical decision support.

To evaluate the CT radiomics models' performance, we trained and compared several models, including Logistic Regression, K-Nearest Neighbors, Support Vector Classifier, XGBoost, Random Forest, and TabPFNv2. Although TabPFNv2 was the novel tabular foundation model [24], however it did not perform best in our experiments. Among these, Random Forest showed the highest diagnostic accuracy. This effectiveness can be attributed to the ensemble nature of Random Forest, which minimizes overfitting risks by constructing each decision tree with different subsets of data and features. Consequently, Random Forest prevents the model from becoming overly specific to the training data while preserving accuracy. Moreover, Random Forest's capacity to handle a large number of features makes it well-suited for capturing the complexity and diversity of radiomic data (such as texture and shape), allowing the model to utilize these features without requiring extensive feature selection [25]. In comparison, other models showed limitations in this context. Logistic Regression and SVC are less capable of modeling complex nonlinear relationships. K-Neighbors is sensitive to high-dimensional noise. XGBoost, although powerful, requires careful hyperparameter tuning. TabPFNv2 [24], as a zero-shot deep learning learner, may not fully leverage domain-specific patterns without task-specific adaptation. These factors make Random Forest more suitable for handling the diversity and complexity of radiomic data in our study.

In this research, the analysis is based on the top 20 most informative radiomic features selected for their diagnostic accuracy in differentiating malignant from benign endometrial tumors (see **Fig.1** and **Fig. 2**). These features, extracted from preoperative CT scans, encompass texture, shape, and intensity characteristics that significantly improve classification accuracy. Multiscale texture features, including first-order statistics and GLCM features derived from "log-sigma" (LoG) scales, capture density variations and subtle intensity differences across different scales [26]. These features are closely associated with the heterogeneous tissue structures characteristic of malignant tumors. First-order features, such as the "Range", offer insights into spatial distribution and intensity variance within lesions. Furthermore, wavelet-decomposed features like the "wavelet-LL_firstorder_Range" and "wavelet-LH_glrlm_RunLengthNonUniformity" enhance the model's ability to detect irregular tissue patterns, further improving its precision in distinguishing benign from malignant structures [27]. This combination of features can serve as AI-driven imaging biomarkers, supporting a non-invasive, nuanced assessment of EC, enabling a more personalized approach to diagnostic and therapeutic planning.

Current intelligent imaging diagnostic models face significant challenges in clinical practice, primarily due to poor interpretability of these models [28,29]. Therefore, enhancing model interpretability is a prerequisite for clinical adoption. In this study, we utilized two visualization techniques, SHAP feature analysis and radiomics feature maps, to assist the model in EC prediction. SHAP values help clinicians understand which features contribute most to each prediction, making the model's decisions more transparent and trustworthy. This integrated explainability approach can also serve as a reference for other studies.

In clinical practice, the early and precise identification of the malignant or benign nature of EC can assist clinicians in patient stratification, as benign and malignant EC patients follow distinct therapeutic pathways. After a patient undergoes CT scanning, this method can automatically extract radiomic biomarkers, generate feature maps, and calculate prediction probabilities. These outputs can then serve as reference information for clinicians to support treatment decision-making, thereby achieving the goal of precision medicine.

Despite the encouraging results, this study has several limitations. The sample size is limited, restricts the model's generalizability as it may not fully capture the diversity of EC presentations. Additionally, the data were sourced from only two hospitals, potentially introducing bias and limiting the model's applicability across different imaging settings, equipment, and healthcare environments. Nevertheless, by adopting strategies such as grouping by hospital center and performing cross-validation, we can preliminarily ensure that our model possesses high predictive accuracy and generalizability. To further enhance robustness, future research could benefit from larger, more diverse datasets, improving the model's generalizability in varied clinical settings. In addition, the incorporation of emerging technologies such as large language models [30,31] and human-in-the-loop [32] approaches may further enhance the potential for deploying the model in clinical practice. Finally, due to the constraints of a retrospective study, we did not include more pathological information. If the selected radiomic features could be further correlated with pathological characteristics, it would further enhance the biological interpretability of the radiomic features [33].

## Conclusion

In this study, we developed and tested ML models based on radiomic features from CT imaging to differentiate between malignant and benign lesions in EC. The results demonstrated that CT-based radiomics analysis achieved high diagnostic accuracy, making it a promising intelligent auxiliary tool for the diagnosis of EC. Notably, the random forest model exhibited the best diagnostic performance. This suggests that combining radiomics from CT imaging with advanced ML algorithms can significantly improve early diagnostic accuracy for EC, thereby aiding clinicians in making more precise decisions and potentially improving patient outcomes.

## Materials and Methods

This study was conducted in accordance with the tenets of the Declaration of Helsinki (as revised in 2013). The study protocol was approved by institutional review boards of Qingdao Hiser Hospital and Qingdao Women and Children Hospital. All participants provided informed consent.

### Patient population

A total of 83 patients diagnosed with benign endometrial conditions were included in this study (**Fig. 5**). The inclusion criteria were as follows: (1) patients with histologically confirmed uterine adenocarcinoma, (2) patients without distant metastasis, (3) patients who were not excluded due to pregnancy. The exclusion criteria were as follows: (1) patients lacking basic clinical information, such as age and sex, (2) those with unclear clinical staging, and (3) those with poor CT image quality. We divided the patients into a training set and an independent testing set based on the hospital where they received treatment, with the proportion being approximately 7:3. The training set consisted of 59 patients from Qingdao Hiser Hospital treated between January 2018 and March 2020, and the testing set included 24 patients from Qingdao Women and Children Hospital treated between November 2018 and March 2020. Clinical characteristics were acquired from all patients, see **Table 1.**

**Pathological Diagnosis**

Pathological diagnosis remains the gold standard for confirming endometrial cancer [34]. This involves histopathological analysis of tissue samples obtained through endometrial biopsy methods, such as diagnostic curettage and hysteroscopy. Diagnostic curettage allows for comprehensive tissue sampling from both the endocervix and endometrial cavity, which aids in assessing the lesion's extent and nature. Hysteroscopy further enhances diagnostic accuracy by providing direct visualization, targeted biopsy, and excision of localized lesions. Histopathological examination of these samples offers critical insights into the characteristics and spread of the disease, which supports the development of personalized treatment plans for endometrial cancer.

**CT image acquisition**

Spiral CT scans for endometrial cancer were performed on all patients (using Philips Brilliance iCT and Siemens Somatom Definition AS). Both scanners are used for both hospitals. The scanning parameters were as follows: tube voltage at 120 kV, automatic tube current, pitch 1.0~1.5, matrix 512 × 512, and field of view (FOV) 350 mm × 350 mm. After initial data collection, all patients underwent a no-interval reconstruction of 0.5 ~ 3.0 mm. A high-resolution algorithm was applied to enhance image quality. The scans were conducted with the patient in a supine position, arms placed on either side of the head, and breath-hold was employed during scanning. The scanning range covered the area from above the uterine fundus to below the pelvic cavity. All CT images were retrieved from the Picture Archiving and Communication System (PACS) for further feature extraction and analysis.

**Tumor segmentation**

Tumor segmentation was manually conducted across the entire tumor volume using ITK-SNAP software (version 3.8.0; www.itksnap.org). Region-of-interest (ROI) positioning was established by two board-certified gynecologic radiologists with 11 and 13 years of experience in endometrial cancer imaging, respectively. Both radiologists were blinded to clinical and histological findings to reduce potential bias. To evaluate the impact of inter-observer variability on ROI delineation, which could affect radiomic feature extraction, each radiologist independently reviewed all CT images. Any discrepancies were resolved by consensus. During the process of determining the final ROIs, we also calculated the inter-observer agreement using the Intraclass Correlation Coefficient (ICC) metric. The results showed ICC values ranging from 0.812 to 0.920, indicating strong consistency in the manual segmentation outcomes.

**Radiomics feature extraction, selection, and analysis**

In this study, radiomic features were extracted from CT images using PyRadiomics (version 3.0.1). A total of 1122 features were obtained from each region of interest (ROI), encompassing both original and transformed image types. From the original image, 121 radiomic features were extracted, including 14 shape features, 18 first-order statistical features, and 89 texture features derived from gray level co-occurrence matrix (GLCM), gray level run length matrix (GLRLM), gray level size zone matrix (GLSZM), gray level dependence matrix (GLDM), and neighboring gray-tone difference matrix (NGTDM).

In addition to the original image, features were also extracted from several derived image types, including eight wavelet-decomposed images, square-transformed images, square root-transformed images, logarithmic-transformed images, exponential-transformed images, and Laplacian of Gaussian (LoG)-filtered images with multiple sigma values. For each of these transformed images, 91 features comprising first-order and texture features were extracted, excluding shape features. The combination of features from the original and transformed images resulted in a comprehensive radiomic profile of 1122 features per ROI. All features were standardized before being included in the subsequent feature selection and modeling process.

The predictive performance of the features was evaluated using ridge regression-based recursive feature elimination (RFE), through which valuable features were selected for modeling. RFE identifies the best features by repeatedly building models and evaluating feature coefficients [23]. To ensure stable estimates, the 20 most representative features were selected from the original 1122 high-throughput radiomic features for modeling. To determine the impact of the number of included features on the optimal model, we also compared the modeling results using 5, 10, and 20 radiomic features, respectively (see Supplementary Figure 2).

Based on the selected features, the relationships among these selected features were first analyzed, followed by a clinical correlation analysis to assess their biological significance. Then, the Shapley Additive Explanations (SHAP) analysis further evaluated the feature importance ranking to enhance model interpretability. Meanwhile, the radiomics feature maps were calculated to further enhance the understanding of feature visualization [27, 35].

**Model construction and validation**

The prediction models were built using ML models including random forest, logistic regression, support vector machine, XGBoost and TabPFNv2 [24]. Accuracy, precision, sensitivity, and specificity metrics were used to evaluate the performance of each model. Finally, after comparing all these models, the final Radiomic score was calculated in the training set and tested in an external testing set. Meanwhile, the receiver operating characteristic (ROC) curve, confusion matrix, calibration curve, and decision curve were all calculated and evaluated.

**Statistical analysis**

All statistical analyses and machine learning algorithms were performed using Python (version 3.10). The evaluation of the model was primarily conducted using the area under the ROC curve (AUROC) and area under the precision-recall curve (AUPRC), with the 95% Confidence Interval (CI). These metrics effectively help us analyze the trade-offs between true positive rates and false positive rates at various thresholds, thus assessing the model's classification effectiveness [36, 37]. We also applied the DeLong test to compare the efficacy differences between models, determining which model exhibited superior statistical performance. All tests had to achieve statistical significance at a P-value less than 0.05 to ensure the reliability and validity of the results. Through these rigorous statistical methods, we ensured that our model is not only theoretically sound but also robust in practical application, may providing strong predictive performance.

**Abbreviations**

EC: Endometrial Cancer

LND: Lymphadenectomy

CT: Computed Tomography

MRI: Magnetic Resonance Imaging

PACS: Picture Archiving and Communication System

---

ROI: Region of Interest

RFE: Recursive Feature Elimination

GLCM: Gray Level Co-occurrence Matrix

GLRLM: Gray Level Run Length Matrix

GLSZM: Gray Level Size Zone Matrix

GLDM: Gray Level Dependence Matrix

NGTDM: Neighboring Gray-Tone Difference Matrix

AUROC: Area Under the Receiver Operating Characteristic Curve

AUPRC: Area Under the Precision-Recall Curve

ROC: Receiver Operating Characteristic

DCA: Decision Curve Analysis

MMR-D: Mismatch Repair Deficiency

TMB-H: Tumor Mutation Burden-High

RF: Random Forest

LR: Logistic Regression

KNeighbor: K-Nearest Neighbors

SVC: Support Vector Classifier

**Ethics approval and consent to participate**

The study was conducted following the tenets of the Declaration of Helsinki. Ethical approval was obtained from the institutional ethics committees of both Qingdao Hiser Hospital and Qingdao Women and Children Hospital. Written informed consent was obtained from all participants prior to inclusion in this study.

**Competing interests**

The authors declare that they have no conflict of interest.

**Funding**

This work was financially supported by the Shandong Province Traditional Chinese Medicine Science and Technology Project (M-2022012) and Chengdu Health Commission Medical Health Project (No.2022667).

**References**

[1] Miller K.D., Nogueira L., Devasia T., Mariotto A.B., Yabroff K.R., Jemal A., Kramer J., Siegel R.L. Cancer treatment and survivorship statistics, 2022. CA Cancer J. Clin. 2022; 72:409–436.

[2] Makker V., MacKay H., Ray-Coquard I., Levine D.A., Westin S.N., Aoki D., Oaknin A. Endometrial cancer. Nat. Rev. Dis. Primers. 2021; 7:88.

[3] Siegel R.L., Miller K.D., Wagle N.S., Jemal A. Cancer statistics, 2023. CA Cancer J. Clin. 2023; 73:17–48.

[4] Sbarra, M., Lupinelli, M., Brook, O. R., Venkatesan, A. M., & Nougaret, S. Imaging of endometrial cancer. Radiologic Clinics of North America. 2023; 61(4):609-625.

[5] Burbos, N., Musonda, P., Giarenis, I., Shiner, A. M., & Morris, E. P. Predicting the risk of endometrial cancer in postmenopausal women presenting with vaginal bleeding: the Norwich DEFAB risk assessment tool. British Journal of Cancer. 2012; 106(9):1611-1616.

[6] Lin MY, Dobrotwir A, McNally O, Abu-Rustum NR, Narayan K. Role of imaging in the routine management of endometrial cancer. Int J Gynaecol Obstet. 2018;143:109–17.

[7] Barretina-Ginesta, M. P., Quindós, M., Alarcón, J. D., Esteban, C., Gaba, L., Gómez, C., Pérez Fidalgo, J. A., Romero, I., Santaballa, A., & Rubio-Pérez, M. J. (2022). SEOM–GEICO clinical guidelines on endometrial cancer. Clinical

Guides in Oncology. 2021; 24:625–634.

[8] Hodneland, E., Andersen, E., Wagner-Larsen, K.S. et al. Impact of MRI radiomic feature normalization for prognostic modelling in uterine endometrial and cervical cancers. Sci Rep. 2024; 14:16826.

[9] Azeroual, S., Ben-Bouazza, Fe., Naqi, A. et al. Predicting disease recurrence in breast cancer patients using machine learning models with clinical and radiomic characteristics: a retrospective study. J Egypt Natl Canc Inst. 2024; 36, 20.

[10] Ji J, Ju S and Cai W. Editorial: Radiomics-based theranostics in cancer precision medicine. Front. Oncol. 2023; 13:1250079.

[11] Barretina-Ginesta, M. P., Quindós, M., Alarcón, J. D., Esteban, C., Gaba, L., Gómez, C., Pérez Fidalgo, J. A., Romero, I., Santaballa, A., & Rubio-Pérez, M. J. SEOM-GEICO clinical guidelines on endometrial cancer. Clinical Guides in Oncology. 2022; 24:625-634.

[12] Sbarra, M., Lupinelli, M., Brook, O. R., Venkatesan, A. M., & Nougaret, S. Imaging of Endometrial Cancer. Radiologic Clinics of North America. 2023; 61(4):609-625.

[13] NCCN Guidelines Updates for Endometrial Cancer Management. Journal of the National Comprehensive Cancer Network. 2024; 22(Suppl):49-50.

[14] Williams, L. E., Zheng, Y., Voss, E. A., & Gillies, R. J. Radiomic features from CT images as predictive biomarkers for endometrial cancer aggressiveness. Academic Radiology. 2022; 29(2):273-282.

[15] Veeraraghavan, H., Lee, S., Karlan, B. Y., & Tavassoli, F. A. Advanced deep learning techniques in MRI-based diagnosis of endometrial cancer. Journal of Magnetic Resonance Imaging. 2021; 53(1):188-197.

[16] Smith, A. B., Jones, J. C., Roberts, J., Lee, D. K., & Challis, B. G. Comparative analysis of machine learning algorithms in the MRI-based diagnosis of endometrial cancer. Radiology. 2020; 295(3):215-224.

[17] Aerts, H. J. W. L. et al. Decoding tumour phenotype by noninvasive imaging using a quantitative radiomics approach. Nature Communications. 2014; 5:4006.

[18] Esteva, A., Kuprel, B., Novoa, R. A., Ko, J., Swetter, S. M., Blau, H. M., & Thrun, S. Dermatologist-level classification of skin cancer with deep neural networks. Nature. 2017; 542(7639):115-118.

[19] Litjens, G., Sánchez, C. I., Timofeeva, N., Hermsen, M., Nagtegaal, I., Kovacs, I., Hulsbergen-Van de Kaa, C., Bult, P., Van Ginneken, B., & Van Der Laak, J. Deep learning as a tool for increased accuracy and efficiency of histopathological diagnosis. Scientific Reports. 2016; 6:26286.

[20] Urban, G., Tripathi, P., Alkayali, T., Mittal, M., Jalali, F., Karnes, W., & Baldi, P. Deep learning localizes and identifies polyps in real time with 96% accuracy in screening colonoscopy. Gastroenterology. 2018; 155(4):1069-1078.e8.

[21] Veeraraghavan, H., Friedman, C.F., DeLair, D.F. et al. Machine learning-based prediction of microsatellite instability and high tumor mutation burden from contrast-enhanced computed tomography in endometrial cancers. Sci Rep. 2020; 10, 17769.

[22] Coada CA, Santoro M, Zybin V, Di Stanislao M, Paolani G, Modolon C, Di Costanzo S, Genovesi L, Tesei M, De Leo A, Ravegnini G, De Biase D, Morganti AG, Lovato L, De Iaco P, Strigari L, Perrone AM. A Radiomic-Based Machine Learning Model Predicts Endometrial Cancer Recurrence Using Preoperative CT Radiomic Features: A Pilot Study. Cancers (Basel). 2023; 15(18):4534.

[23] Wang L, Wen D, Yin Y, et al. Musculoskeletal ultrasound image - based radiomics for the diagnosis of Achilles tendinopathy in skiers. Journal of

Ultrasound in Medicine. 2023, 42(2): 363-371.

[24] Hollmann N, Müller S, Purucker L, et al. Accurate predictions on small data with a tabular foundation model. Nature. 2025; 637(8045): 319-326.

[25] Jiang Z, Wang B, Han X, Zhao P, Gao M, Zhang Y, Wei P, Lan C, Liu Y, Li D. Multimodality MRI-based radiomics approach to predict the posttreatment response of lung cancer brain metastases to gamma knife radiosurgery. Eur Radiol. 2022; 32(4):2266-2276.

[26] Dong Y, Jiang Z, Li C, Dong S, Zhang S, Lv Y, Sun F, Liu S. Development and validation of novel radiomics-based nomograms for the prediction of *EGFR* mutations and Ki-67 proliferation index in non-small cell lung cancer. Quant Imaging Med Surg. 2022;12(5):2658-2671.

[27] Jiang Z, Yin J, Han P, Chen N, Kang Q, Qiu Y, Li Y, Lao Q, Sun M, Yang D, Huang S, Qiu J, Li K. Wavelet transformation can enhance computed tomography texture features: a multicenter radiomics study for grade assessment of COVID-19 pulmonary lesions. Quant Imaging Med Surg. 2022;12(10):4758-4770.

[28] Raj G M, Dananjayan S, Gudivada K K. Applications of artificial intelligence and machine learning in clinical medicine: What lies ahead?. Med Adv, 2024, 2(2): 202-204.

[29] Yuan H, Yu K, Xie F, et al. Automated machine learning with interpretation: A systematic review of methodologies and applications in healthcare. Medicine Advances, 2024, 2(3): 205-237.

[30] Ali H. Generative pre - trained transformer 4 in healthcare: challenges, opportunities, and recommendations. Medicine Advances, 2023, 1(2): 163-166.

[31] Jiang Z, Cheng D, Qin Z, et al. TV-SAM: increasing Zero-Shot segmentation performance on multimodal medical images using GPT-4 generated descriptive prompts without human annotation. Big Data Mining and Analytics, 2024, 7(4): 1199-1211.

[32] Yuan H, Kang L, Li Y, et al. Human - in - the - loop machine learning for healthcare: current progress and future opportunities in electronic health records. Medicine Advances, 2024, 2(3): 318-322.

[33] Jiang C, Qian C, Jiang Q, et al. Virtual biopsy for non-invasive identification of follicular lymphoma histologic transformation using radiomics-based imaging biomarker from PET/CT. BMC medicine, 2025, 23(1): 49.

[34] Morice P, Leary A, Creutzberg C, et al. Endometrial cancer. The Lancet. 2016; 387(10023): 1094-1108.

[35] Xie J, Yang Y, Jiang Z, et al. MRI radiomics-based decision support tool for a personalized classification of cervical disc degeneration: a two-center study. Frontiers in Physiology. 2024; 14:1281506.

[36] Ye, R., Jiang, Z., Shao, R., Yan, Q., Zhou, L., Zhang, T., & Sun, Y. Development and validation of tongue imaging-based radiomics tool for the diagnosis of insomnia degree: a two-center study. Medical Data Mining. 2024;7(1):4.

[37] Carrington A M, Manuel D G, Fieguth P W, et al. Deep ROC analysis and AUC as balanced average accuracy, for improved classifier selection, audit and explanation. IEEE Transactions on Pattern Analysis and Machine Intelligence. 2022; 45(1): 329-341.

**Tables**

**Table 1** Patient demographic characteristics

| Characteristics | Groups | Full Cohort (n=83) | Training Set (n=59) | Testing Set (n=24) | *P* |
|---|---|---|---|---|---|
| Age (years) | Mean ± SD | 57.46 8.99 | 56.66 8.86 | 59.42 9.20 | 0.218 |
| Gender | Male | 0 | 0 | 0 | 1.000 |
| | Female | 83 | 59 | 24 | |
| Diagnosis Type | Benign | 37 | 26 | 11 | 0.886 |
| | Malignant | 46 | 33 | 13 | |

**Table 2** Top 20 radiomic features comparison of Malignant and Benign groups

| Num | Radiomic Features | Malignant Group Mean ± SD | Benign Group Mean ± SD | *P* |
|---|---|---|---|---|
| F1 | LoG-5-0-mm-firstorder_90Percentile | 0.8973 ± 2.7674 | -0.9706 ± 0.7182 | <0.001 |
| F2 | wavelet-LH_firstorder_Maximum | 33.2488 ± 33.5727 | 14.9921 ± 4.8543 | 0.003 |
| F3 | squareroot_glcm_Idn | 0.9718 ± 0.0198 | 0.9596 ± 0.0121 | 0.005 |
| F4 | LoG-1-0-mm-glcm_MaximumProbability | 0.3875 ± 0.0577 | 0.4283 ± 0.0250 | 0.001 |
| F5 | squareroot_gldm_DependenceNonUniformity | 16142.3282 ± 11937.4632 | 8633.8018 ± 6180.8670 | 0.003 |
| F6 | wavelet-HL_firstorder_Maximum | 28.4411 ± 25.4365 | 13.6784 ± 9.4997 | 0.003 |
| F7 | LoG-3-0-mm-glcm_MaximumProbability | 0.5927 ± 0.1446 | 0.6951 ± 0.0667 | 0.001 |
| F8 | log-sigma-3-0-mm-3D_glrlm_GrayLevelNonUniformity | 4232.4170 ± 3772.8624 | 2097.2949 ± 1207.3108 | 0.003 |
| F9 | wavelet-LH_glrlm_RunLengthNonUniformity | 26007.2400 ± 16713.7003 | 14698.5115 ± 10658.1750 | 0.002 |
| F10 | LoG-1-0-mm-firstorder_Maximum | 28.3740 ± 28.7366 | 11.7850 ± 4.5883 | 0.002 |
| F11 | log-sigma-3-0-mm-3D_glcm_JointEnergy | 0.4245 ± 0.1408 | 0.5147 ± 0.0848 | 0.003 |
| F12 | original_glcm_MaximumProbability | 19854.2579 ± 14212.2986 | 12160.5326 ± 4979.3845 | 0.005 |
| F13 | wavelet-LH_firstorder_Range | 71.2081 ± 65.8776 | 33.7422 ± 10.9061 | 0.002 |
| F14 | wavelet-HL_glrlm_RunLengthNonUniformity | 19522.2754 ± 14082.0266 | 10815.4521 ± 7310.8897 | 0.003 |
| F15 | original_firstorder_Range | 112.8110 ± 92.6035 | 56.7198 ± 24.6335 | 0.001 |
| F16 | wavelet-LL_firstorder_Range | 211.0077 ± 173.9878 | 104.9770 ± 50.7249 | 0.001 |
| F17 | log-sigma-1-0-mm-3D_glcm_JointEnergy | 0.2717 ± 0.0507 | 0.3003 ± 0.0148 | 0.003 |
| F18 | log-sigma-1-0-mm-3D_glrlm_ShortRunEmphasis | 0.3876 ± 0.1447 | 0.3087 ± 0.0430 | 0.004 |
| F19 | wavelet-LH_ngtdm_Complexity | 96.3095 ± 142.3560 | 18.9783 ± 12.6605 | 0.003 |
| F20 | log-sigma-3-0-mm-3D_firstorder_Uniformity | 0.4881 ± 0.1366 | 0.5699 ± 0.0797 | 0.005 |

**Table 3** Performance comparison of various machine learning models for classification.

| **Model** | **Dataset** | **AUROC (95%CI)** | **Sensitivity (%)** | **Specificity (%)** | **Accuracy (%)** | **Precision (%)** | **AUPRC** |
|---|---|---|---|---|---|---|---|
| Logistic Regression | Training | 0.95[0.88,0.99] | 88.57 | 87.50 | 88.14 | 91.18 | 0.97 |
| | Testing | 0.95[0.87,1.00] | 100.00 | 92.31 | 95.83 | 91.67 | 0.95 |
| K-Neighbors | Training | 0.94[0.88,0.99] | 82.86 | 91.67 | 86.44 | 93.55 | 0.94 |
| | Testing | 0.94[0.83,1.00] | 81.82 | 92.31 | 87.5 | 90.00 | 0.89 |
| SVC | Training | 0.97[0.91,1.00] | 94.29 | 91.67 | 93.22 | 94.29 | 0.98 |
| | Testing | 0.95[0.83,1.00] | 100.00 | 92.31 | 95.83 | 91.67 | 0.93 |
| XGBoost | Training | 1.00[1.00,1.00] | 100.00 | 100.00 | 100.00 | 100.00 | 1.00 |
| | Testing | 0.88[0.71,1.00] | 72.73 | 92.31 | 83.33 | 88.89 | 0.84 |
| TabPFNv2 | Training | 1.00[1.00,1.00] | 100.00 | 100.00 | 100.00 | 100.00 | 1.00 |
| | Testing | 0.96[0.85,1.00] | 81.82 | 92.31 | 87.50 | 90.00 | 0.94 |
| **Random Forest** | **Training** | **1.00[1.00,1.00]** | **100.00** | **95.83** | **98.31** | **97.22** | **1.00** |
| | **Testing** | **0.96[0.89,1.00]** | **100.00** | **92.31** | **95.83** | **91.67** | **0.94** |

## Figures

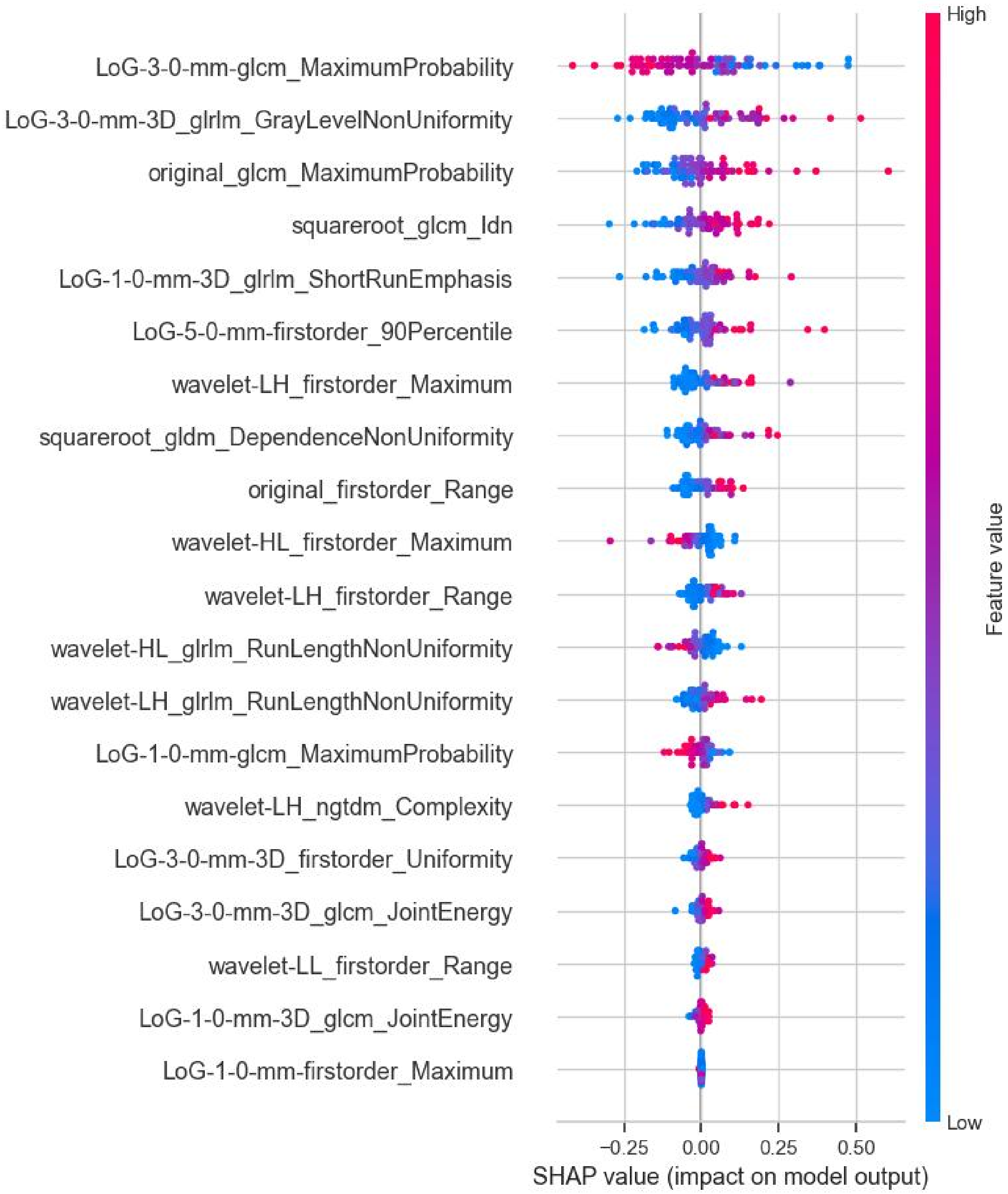

**Fig. 1** SHAP analysis of the Top 20 radiomic features.

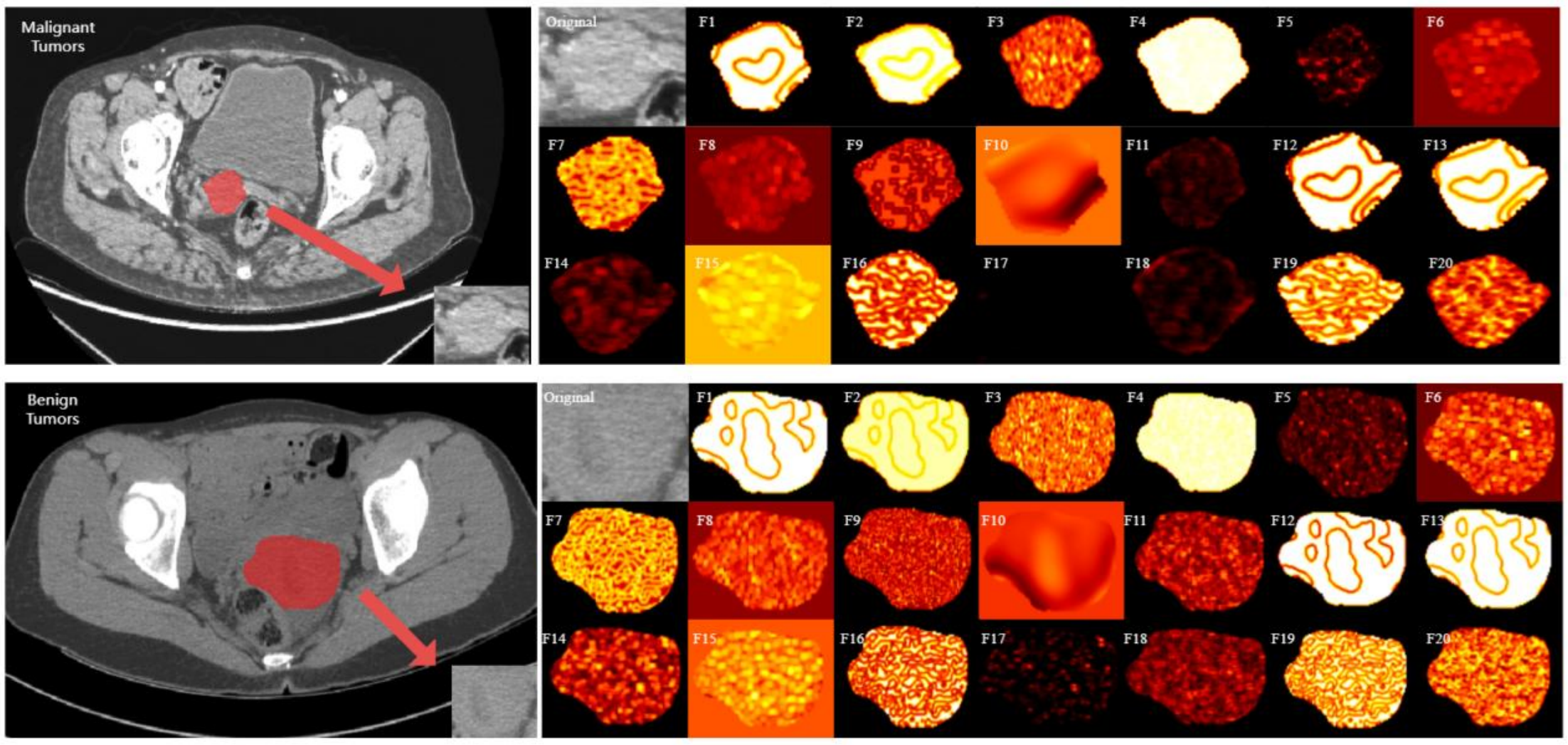

**Fig. 2** Two representative cases demonstrating the precise prediction of EC malignancy and benignity with our radiomics feature maps.

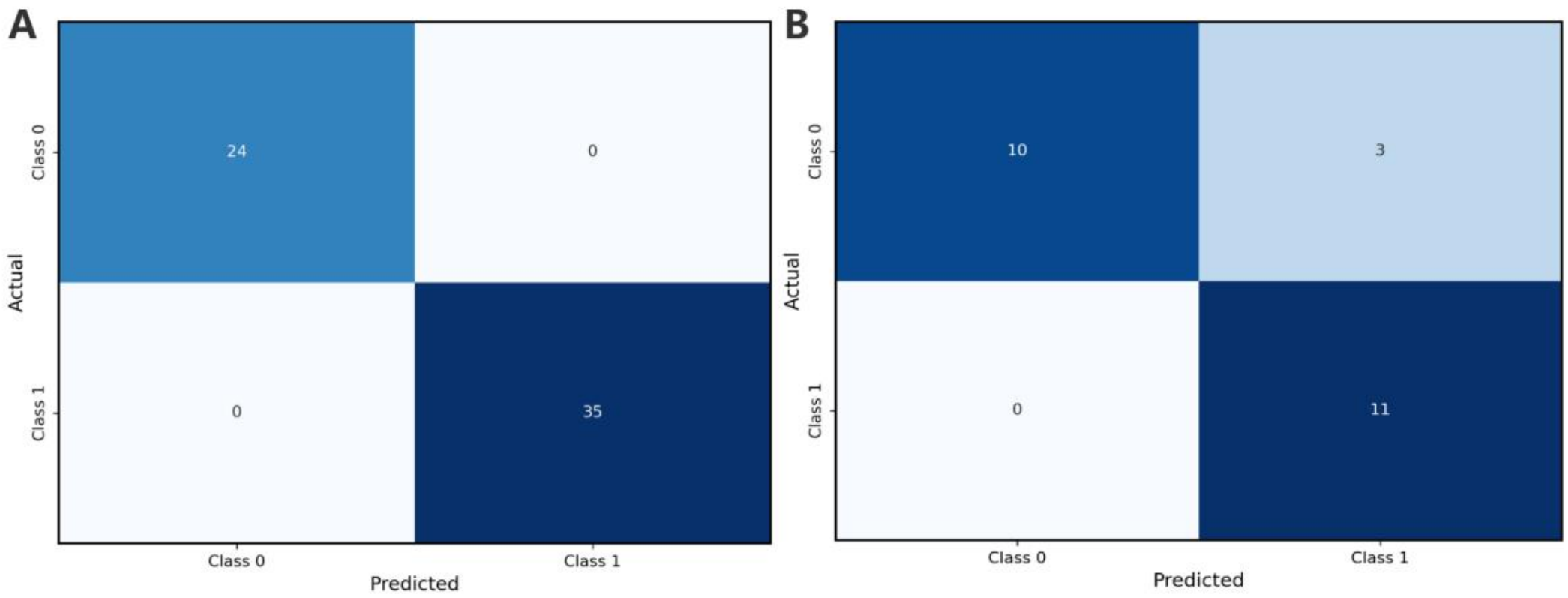

**Fig. 3** Confusion matrices of the best ML model. (A) Training set confusion matrix for the best model. (B) Testing set confusion matrix for the best model.

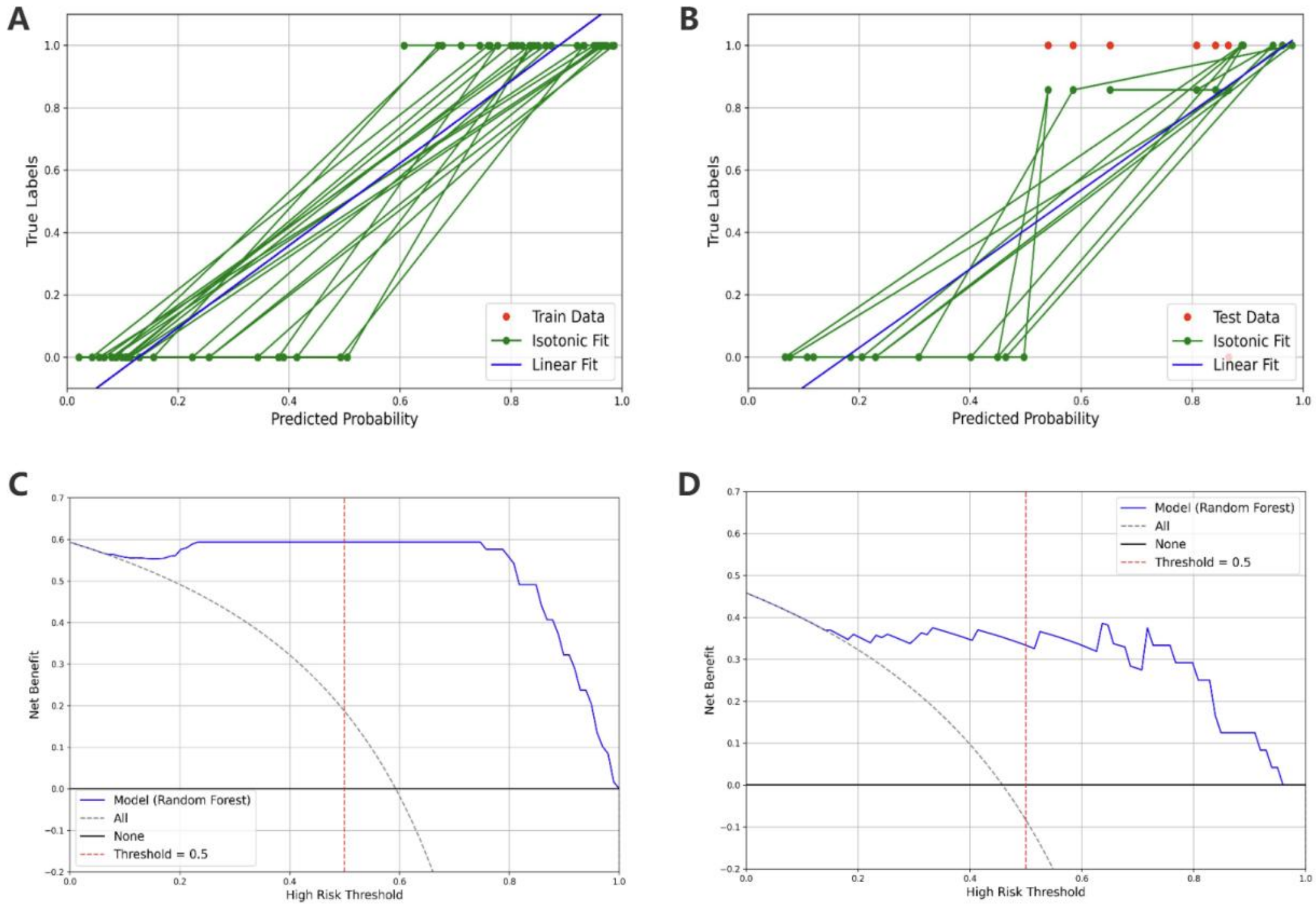

**Fig. 4** Calibration curves and decision curves for the final model. Calibration curves in (A) training set and (B) testing set. Decision curves in the (C) training set and (D) testing set.

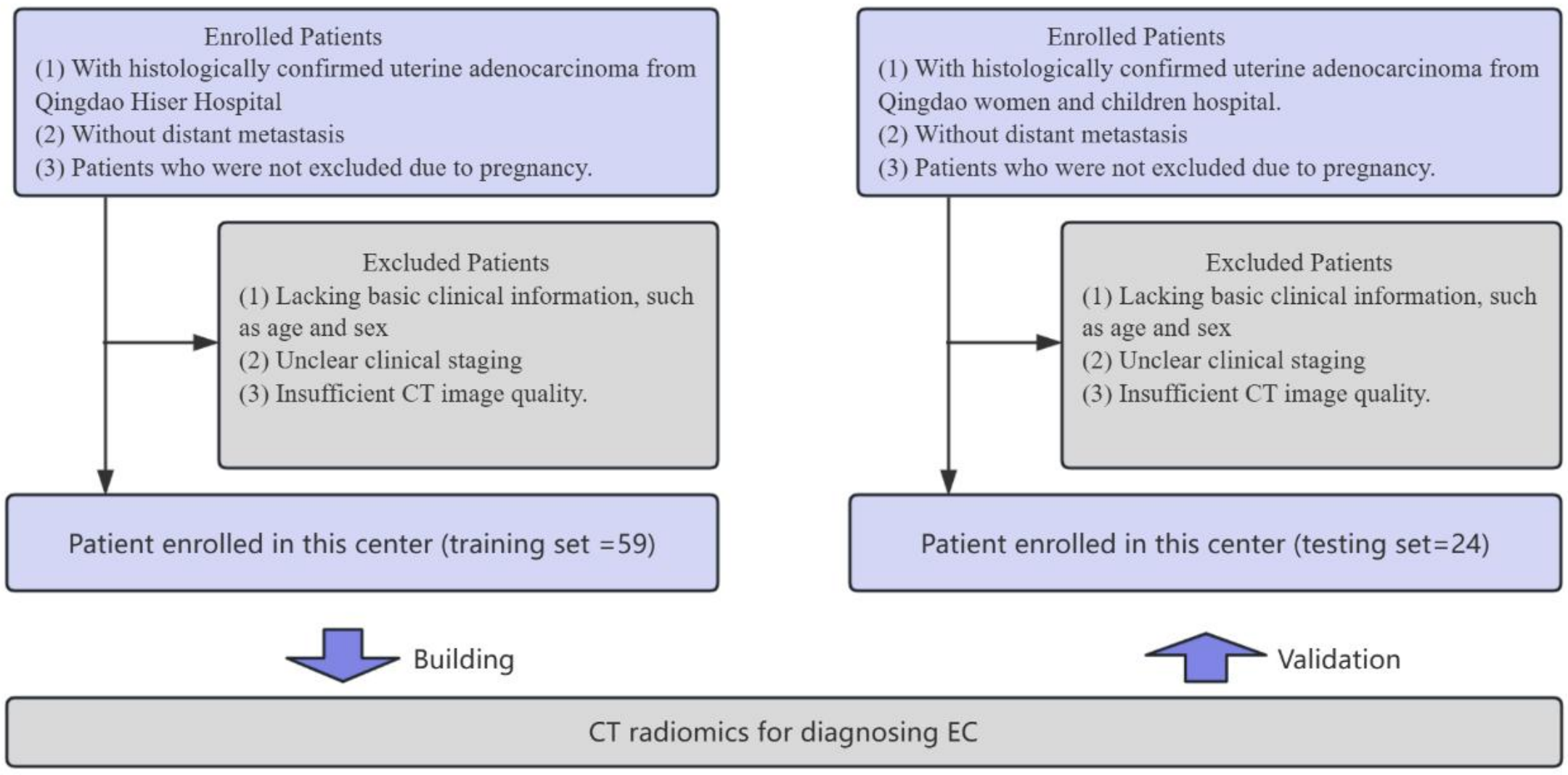

**Fig.5** Patient inclusion and exclusion criteria flowchart.